\documentclass[preprint,aps,prb]{revtex4}

\usepackage{amsmath}

\begin{document}

\title{Generalization of the Schott energy in electrodynamic radiation
theory}

\author{Jos\'{e} A. Heras}

\affiliation{Departamento de F\'isica, E. S. F. M., Instituto
Polit\'ecnico Nacional, A. P. 21-081, 04021, M\'exico D. F., M\'exico and
Department of Physics and Astronomy, Louisiana State University,
Baton Rouge, Louisiana 70803-4001, USA}

\author{R. F. O'Connell}

\affiliation{Department of Physics and Astronomy, Louisiana State
University,
Baton Rouge, Louisiana 70803-4001, USA}


\begin{abstract}
We discuss the origin of the Schott energy in the
Abraham-Lorentz version of electrodynamic radiation theory and how it
can be used to explain some apparent paradoxes. We also derive the
generalization of this quantity for the Ford-O'Connell equation,
which has the merit of being derived exactly from a
microscopic Hamiltonian for an electron with structure and has been shown
to be free of the problems associated with the Abraham-Lorentz theory.
We emphasize that the instantaneous power supplied by the applied force
not only gives rise to radiation (acceleration fields), but it can change
the kinetic energy of the electron and change the Schott energy
of the velocity fields. The important role played by boundary conditions
is noted.
\end{abstract}

\maketitle

\section{INTRODUCTION}

Since its introduction in 1912,\cite{schott} there has been much confusion 
concerning the nature of the Schott energy, despite the fact that some
authors\cite{gin,cole,grif} have presented a clear explanation of its origin.
It arises from the fact that the power supplied by an external force to a
charged particle not only contributes to the energy radiated
(acceleration fields) but also to the velocity fields.  This feature is
not connected with the well-known deficiencies of the Abraham-Lorentz
theory (runaway solutions, etc.). Previous discussions of the Schott
energy arose in the context of the Abraham-Lorentz equation of motion for
a radiating electron. In this paper we  define a (generalized) Schott
energy that is applicable not only to the Abraham-Lorentz theory but to
all theories of a radiating electron. To do so, we start by recalling the
classical Newtonian equation of motion for a particle of mass
$M$ under the action of an external field
$\vec{f}(t)$:
\begin{equation}
M\vec{a}=\vec{f},
\end{equation}
where $\vec{a}$ is the acceleration. 
The only effect of $f(t)$ is to change the kinetic energy $T$ of the
particle, where
\begin{equation}
T=\frac{1}{2}M{v^{2}},
\end{equation}
and $v$ is the velocity. We stress these elementary
facts because they are often overlooked in the development of radiation
theories for a charged particle because for $\vec{f}(t)=0$,
the equation of motion should reduce to $M\vec{a}=0$. In
particular, this requirement is not obeyed by the Abraham-Lorentz equation
(so that runaway solutions emerge) whereas it is obeyed by the Ford-O'Connell
theory as we will discuss.

We consider an electron of charge $e$ subject to a
external force $\vec{f}(t)$. The total work done by $\vec{f}(t)$
during an arbitrary time interval consists of three parts
(1) the change in kinetic energy $\Delta{T}$ (which is
independent of $e$), 
(2) the radiated energy, which is the energy in the acceleration
or far fields,\cite{jackson} and
(3) the change in energy in the velocity or near fields,\cite{jackson}
which does not give rise to radiation. This change can be positive or
negative.

The energy in the velocity fields is the Schott energy. Thus,
at any time $t$, the instantaneous power, $P(t)$, supplied by
$\vec{f}(t)$ does not contribute just to the radiated energy.
A feature of the Schott energy is that its time derivative appears in the
expression for $P(t)$ so that ``\ldots if we\ldots consider only intervals
over which the system returns to its initial state, then the energy in the
velocity fields is the same at both ends, and the only net loss is in the
form of radiation\ldots''\cite{grif} That is, the Schott energy is the
energy contributed to the velocity fields by the external field and does
not contribute to the radiated energy (which is due to acceleration fields).

A related question is whether or not
radiation can occur for constant acceleration (because the Larmor result
for the radiated energy depends only on the acceleration squared whereas
the radiation reaction term in the Abraham-Lorentz equation of motion
depends on the rate of change of the acceleration). The solution to this
apparent paradox is summarized succintly in Ref.~\onlinecite{gin} where it
is noted that ``\ldots\ the radiated energy and the work of the radiation
friction are not equal to each other in the nonstationary state,'' due again
to the existence of the Schott energy. For that reason, it is desirable to
consider energy exchange between the particle and the field at each instant
of time, rather than using conservation laws integrated over time.

There has also been a long-standing recognition
that the Abraham-Lorentz analysis has a fundamental flaw related to the
existence of runaway solutions, which are a manifestation of the fact that
causality is violated.\cite{jackson} A solution to the latter problem was
presented by Ford and O'Connell\cite{ford1,ford2,ford3,oconnell} who
pointed out the necessity of ascribing structure to the electron. Their
solution led to a second-order equation of motion that is
simple and well-behaved and incorporates quantum and fluctuation effects and
the presence of a potential $V$. Here we confine ourselves to the
non-relativistic classical case with $V=0$, which is the case most often
considered in the literature.\cite{grif,jackson}

In Sec.~II we consider the generalization of the Schott energy for the
radiation reaction force $F_{d}$ (the subscript $d$ indicates its
dissipative nature) without specifying its specific form. A key feature of
our analysis is that because the electron motion and the rate of radiation
are continually changing in time, we consider conservation of power (the
power $P(t)$ supplied by the external force to the particle is equal to
the rate of change of the particle's kinetic energy plus the rate of
change of the velocity fields and the acceleration fields), as distinct
from energy (which is integrated power). The latter gives less
information and obscures the nonstationary aspect of the problem. As a
result, we find that although the radiated power depends on
$d\vec{f}/dt$, the integrated radiated power, that is, the radiated energy,
does not. In Secs.~III and IV we apply our general analysis to the
Abraham-Lorentz and Ford-O'Connell theories and describe the physical
nature of the generalized Schott terms. We also emphasize why the
Ford-O'Connell theory is superior to the Abraham-Lorentz theory. Our
conclusions are presented in Sec.~V.

\section{General Equation of Motion for a Radiating Electron}

The equation of a radiating electron may be written in the form
\begin{equation}
M\vec{a}=\vec{f}+\vec{F_{d}},
\end{equation}
where $\vec{a}$ is the acceleration, $\vec{f}$ is the applied
force, and
$\vec{F_{d}}$ is a dissipative force arising from the back reaction due
to the emitted radiation. All of these quantities are functions of time. The
instantaneous power supplied by the external force $\vec{f}(t)$ to the
electron is
\begin{equation}
P(t)=\vec{f}(t) \cdot \vec{v}(t),
\end{equation}
where $\vec{v}(t)$ is the velocity. From Eq.~(3), we
obtain
\begin{subequations}
\begin{eqnarray}
P(t) &=& M\vec{a} \cdot \vec{v} - \vec{v} \cdot \vec{F_{d}} \\ &=&
\frac{dT}{dt} - \vec{v}\cdot \vec{F_{d}} 
\equiv P_{N}(t)+P_{d}(t),
\end{eqnarray}
\end{subequations}
where the kinetic energy $T$ is given by Eq.~(2). The rate
of change of the kinetic energy of the electron, 
$P_{N}(t)$, arises from the application of Newton's law when
$F_{d}=0$. Our main interest is in the
$P_{d}(t)$ term and as we will discuss, this term contributes not
only to the radiated electromagnetic energy but also to the energy
in the velocity fields.

The total work done by the external
force during the time interval $t_{2}-t_{1}$ is
\begin{subequations}
\begin{eqnarray}
W\equiv{W}(t_{1},t_{2}) &=&
\!\int_{t_{1}}^{t_{2}} P(t^\prime) dt^\prime
\\ &=&
\!\int_{t_{1}}^{t_{2}} P_{d}(t^\prime) dt^\prime+\Delta{T} 
\equiv {W_{d}}+\Delta{T},
\end{eqnarray}
\end{subequations}
where $\Delta{T}=T(t_{2})-T(t_{1})$ is the change in the kinetic energy.
Thus $P_{d}(t)$ is the instantaneous power delivered to the fields by the
external force (only part of which goes into radiated energy); when
$P_{d}(t)=0$ there is no radiated energy. Note that $W_{d}$ is the total
integrated energy transmitted to the fields.

It is useful to write
\begin{equation}
W_{d}=W_{v}+W_{R},
\end{equation}
where from Eqs.~(5) and (6),
\begin{equation}
W_{d}=\!\int_{t_{1}}^{t_{2}} P_{d}(t^\prime) dt^\prime=
-\!\int_{t_{1}}^{t_{2}}\vec{v} (t^\prime)
\cdot\vec{F}_{d}(t^\prime){d}t^\prime,
\end{equation}
$W_{v}$ is the work done on the velocity fields, and
$W_{R}$ is the radiated energy (associated with the acceleration
fields). In other words, the total work done by the external field on
the electron at time $t$ consists of three parts. The total work changes
the kinetic energy of the electron and concomitantly contributes
both to the acceleration fields (which gives rise to radiation) and the
velocity fields (which do not give rise to radiation).

\section{Abraham-Lorentz Theory}

The Abraham-Lorentz equation of motion\cite{jackson} gives
\begin{equation}
\vec{F_{d}}=M\tau\frac{d\vec{a}}{dt},
\end{equation}
so that Eq.~(3) reduces to
\begin{equation}
M\vec{a}=\vec{f}+M\tau\frac{d\vec{a}}{dt},
\end{equation}
where
$\tau=2 e^2/(3Mc^3)=6 \times 10^{-24}$\,s
is proportional to the time it takes light to travel the classical radius
of the electron. We see that when the
acceleration is constant in the Abraham-Lorentz
theory, 
$\vec{F}_{d}$ is zero and thus from Eq.~(5) there is no radiated
energy. More generally, from Eqs.~(5) and (9) we obtain
\begin{subequations}
\begin{eqnarray}
P_{d}(t) &=& -M\tau\vec{v} \cdot \frac{d\vec{a}}{dt} \\
&=&
-M\tau\Big[\frac{d}{dt}(\vec{v} \cdot \vec{a})-a^{2}\Big] 
= P_{L}-\frac{d}{dt}E_{s},
\end{eqnarray}
\end{subequations}
where
\begin{equation}
P_{L}=M\tau{a^{2}},
\end{equation}
is the familiar Larmor rate of radiation,\cite{jackson}
and
\begin{equation}
E_{s}=M\tau(\vec{v} \cdot \vec{a}),
\end{equation}
is the Schott energy. Note that the total time derivative of $E_{s}$ 
appears in the expression for
$P_{d}(t)$. It follows that
\begin{subequations}
\begin{eqnarray}
W_{d} &=& \!\int_{t_{1}}^{t_{2}} P_{L} dt -
\{E_{s}(t_{2})-E_{s}(t_{1})\}
\\ &=&
\!\int_{t_{1}}^{t_{2}} P_{L} dt-M\tau\{\vec{v}(t_{2}) \cdot
\vec{a}(t_{2})-\vec{v}(t_{1})
\cdot \vec{a}(t_{1})\}.
\end{eqnarray}
\end{subequations}
Thus, if the accelerations are equal at times $t_{2}$ and
$t_{1}$ then $W_{v}=0$ and $W_{d}=W_{R}$, the
usual result for the radiated energy. Because the initial and find
velocities are generally different, we see from Eq.~(6) that
$W=W_{R}+\Delta{T}$. The same scenario approximately occurs when
$t_{1}$ and
$t_{2}$ correspond to the times at which the applied force is zero and thus 
from Eqs. (3) and (9) the acceleration is of order
$\tau$ and hence very small.

In addition, the Abraham-Lorentz equation (10) has serious problems. In
particular, when $\vec{f}=0$, it is clear that Eq.~(10) does not reduce to
Newton's equation as it should, and consequently the well-known runaway
solutions emerge. We now turn to the Ford-O'Connell theory which does not
manifest this problem.

\section{Ford-O'Connell Theory}

The Ford-O'Connell theory\cite{ford1,ford2,ford3,oconnell} is based on a
rigorous microscopic approach whose starting 
point is the universally accepted Hamiltonian of
nonrelativistic quantum electrodynamics generalized to allow
for electron structure.\cite{ford4,ford5} The use of Heisenberg's
equation of motion (or the corresponding Poisson equations of motion in
the classical case) leads to an equation of motion that incorporates
electron structure and quantum effects and an arbitrary potential
$V$.\cite{ford1} In the classical limit
and for $V=0$, the Ford-O'Connell equation of motion reduces to the
Abraham-Lorentz equation in the limit of a point particle (and thus, as a
bonus, we have the first Hamiltonian derivation of the Abraham-Lorentz
equation). More generally, electron structure is taken into
account by incorporating a form factor\cite{ford4,ford1} (the
Fourier transform of the charge distribution), which is written in
terms of a large cutoff frequency $\Omega$. The point electron limit
corresponds to
$\Omega\rightarrow\infty$. More generally, small $\Omega$
implies an extended electron structure. As 
shown in Ref.~\onlinecite{ford1}, values of $\Omega$ larger than
$\tau_{e}^{-1}$ lead to violation of causality. This
violation shows that the problem with the Abraham-Lorentz
theory arises from the assumption of a point electron. In addition, choosing
$\Omega=\tau_{e}^{-1}$ (corresponding to the maximum allowed value of
$\Omega$ and hence to the smallest electron structure consistent with
causality), leads in the classical limit and for $V=0$ to
\begin{equation}
\vec{F_{d}}=\tau\frac{d\vec{f}}{dt},
\end{equation}
so that Eq.~(1) becomes
\begin{equation}
M\vec{a}=\vec{f}+\tau\frac{d\vec{f}}{dt}. \label{eq:17}
\end{equation}
Note that $\vec{F}_{d}$
depends on both the electron (through the
factor $\tau$) and the external field. This dependence contrasts with the
corresponding result given by Eq.~(9) for the Abraham-Lorentz theory, where
the external force does not appear explicitly. Also in the Ford-O'Connell
theory, when the applied force $\vec{f}(t)$ is constant,
$\vec{F}_{d}$ is zero and thus from Eq.~(3) we see that there is no
radiation. It also follows that we can write
\begin{subequations}
\begin{eqnarray}
P_{d}(t) &=& -\tau(\vec{v} \cdot \frac{d\vec{f}}{dt})
=
-\tau\Big[\frac{d}{dt}(\vec{v} \cdot \vec{f})-\vec{f} \cdot \vec{a}\Big]
\\ &=&
-\tau\frac{d}{dt}(\vec{v} \cdot
\vec{f})+\frac{\tau}{M}\Big[f^{2}+\frac{\tau}{2}
\frac{d}{dt}f^{2}\Big] \\ &=& P_{\rm FO}-\frac{d}{dt}E_{\rm FO},
\end{eqnarray}
\end{subequations}
where
\begin{equation}
P_{\rm FO}=\frac{\tau}{M}f^{2}, \label{eq:fo}
\end{equation}
is the result obtained in Refs.~\onlinecite{ford2} and \onlinecite{ford3}
for the rate of radiation. In fact, Ford-O'Connell used two different
derivations in obtaining Eq.~(\ref{eq:fo}), one based on energy
conservation\cite{ford2} and another based on a generalization of Larmor's
derivation to include electron structure.\cite{ford3} Also
\begin{equation}
E_{\rm FO}=\tau(\vec{v} \cdot \vec{f})-\frac{\tau^{2}}{2M}f^{2}
\label{eq:efo}
\end{equation}
is the generalization of the Schott energy. It 
follows that the negative of the time derivative of the Schott energy is the
power fed into the velocity fields by the external force. It 
immediately follows that the integrated power is given by
\begin{equation}
W_{d}=\!\int_{t_{1}}^{t_{2}}P_{\rm FO} dt-[E_{\rm FO}(t_{2})-E_{\rm
FO}(t_{1})]. \label{eq:21}
\end{equation}
We note that $E_{\rm FO}$ differs from $E_{s}$ by
terms of order $\tau^{2}$ and that $E_{\rm FO}$ also appears as a total time
derivative in the expression for the instantaneous power radiated. In
contrast to $E_{s}$, $E_{\rm FO}$ vanishes exactly when the applied force is
turned on and off (a more
physically appealing boundary condition than in the Abraham-Lorentz
analysis), in which case $W_{v}=0$ and $W_{d}$ is equal to the first term in
Eq.~(\ref{eq:21}), which is the result obtained in
Refs.~\onlinecite{ford2} and \onlinecite{ford3}. Thus, for a constant
external field,
$P_{d}$ is always zero except when the field is turned on and off, and it is
then that energy is radiated with an average rate given by
Eq.~(\ref{eq:fo}). We point out that Eq.~(\ref{eq:fo}) was obtained in
Ref.~\onlinecite{ford2} by integrating the equation of motion (\ref{eq:17})
and then using energy conservation. The same result was verified in
Ref.~\onlinecite{ford3} by generalizing Larmor's radiation theory to
incorporate electron structure.

\section{Conclusion}

The Schott energy and its generalization corresponds to energy given to
or taken from the velocity fields and always occurs as a total time
derivative in the expression for the instantaneous power supplied by the
external force. Thus the total work done by the applied force is only
equal to the radiated energy plus the change in kinetic energy when
the boundary conditions ensure that the change in the
Schott energy (the energy of the velocity fields) is equal to zero during
the time interval of interest. These
conditions occur naturally in the Ford-O'Connell theory (as distinct from
the Abraham-Lorentz theory) because
$f(t)$ is zero at the initial and final times. Moreover, it is immediately
clear from the Ford-O'Connell equation of motion (\ref{eq:17}), that when
$f=\mbox{constant}$, Eq.~(\ref{eq:17}) reduces to the Newtonian equation of
motion (1). In other words, there is no radiation reaction term in the
equation of motion reflecting the fact that there is no radiation when
$f=\mbox{constant}$, a conclusion that also emerges from a relativistic
generalization.\cite{ford6}  This result is also consistent with
the conclusion\cite{ford7} that an oscillator moving under a
constant force with respect to zero-temperature vacuum does not
radiate despite the fact that it thermalizes at the Unruh
temperature.  Finally, we remark that when quantum effects are taken into
account there are additional fluctuating force terms in the equation of
motion.\cite{ford8}

\begin{acknowledgments}

JAH expresses his gratitude to the Fulbright Program for the scholarship
granted to him to work as a visiting professor in the Department of
Physics and Astronomy, Louisiana State University. He is also grateful
to Louisiana State University for its hospitality.
\end{acknowledgments}

\end{document}